\documentstyle[preprint,aps]{revtex}
\begin{document}
\draft
\title{Reissner-Nordstr$\ddot{o}$m Metric in the Friedman-Robertson-Walker Universe}
\author{Chang Jun Gao$^{1}$\thanks{E-mail: gaocj@mail.tsinghua.edu.cn}
Shuang Nan Zhang$^{1,2,3,4}$\thanks{E-mail:
zhangsn@mail.tsinghua.edu.cn}}
\address{$^{1}$Department of Physics and Center for Astrophysics, Tsinghua University, Beijing 100084, China(mailaddress)}
\address{$^2$Physics Department, University of Alabama in Huntsville, AL 35899, USA }
\address{$^3$Space Science Laboratory, NASA Marshall Space Flight Center, SD50, Huntsville, AL 35812, USA }
\address{$^4$Laboratory for Particle Astrophysics, Institute of High Energy Physics, Chinese Academy of Sciences, Beijing 100039, China}

\date{\today}
\maketitle
\begin{abstract}
\hspace*{7.5mm}The metric for a Reissner-Nordstr$\ddot{o}$m black
hole in the background of the Friedman-Robertson-Walker universe
is obtained. Then we verified it and discussed the influence of
the evolution of the universe on the size of the black hole. To
study the problem of the orbits of a planet in the expanding
universe, we rewrote the metric in the Schwarzschild coordinates
system and deduced the equation of motion for a planet.
\end{abstract}
\section{introduction}
 \hspace*{7.5mm}Black holes
have been investigated in great depth and detail for more than
forty years. However, almost all previous studies have focused on
isolated black holes. On the other hand, one cannot rule out the
important and more realistic situation in which black holes are
actually embedded in the background of universe. Therefore black
holes in non-flat backgrounds form an important topic. \\
\hspace*{7.5mm}As early as in 1933, McVittie [1] found his
celebrated metric for a mass-particle in the expanding universe.
This metric gives us an concrete example for a black hole in the
non-flat background. It is just the Schwarzschild black hole which
is embedded in the Friedman-Robertson-Walker universe although
there was no the notion of black hole at that time. In 1993, the
multi-black hole solution in the background of de Sitter universe
was discovered by Kastor and Traschen [2]. The Kastor-Traschen
solution describes the dynamical system of arbitrary number of
extreme Reissner-Nordstrom black holes in the background of de
Sitter universe. In 1999, Shiromizu and Gen extended it into the
spinning version [3]. In 2000, Nayak etc. [4,5] studied the
solutions for the Schwarzschild and Kerr black holes in the
background of the
Einstein universe.\\
\hspace*{7.5mm}In this letter, we extend the McVittie's solution
into charged black holes. We first deduce the metric for a
Reissner-Nordstr$\ddot{o}$m black hole in the expanding universe;
several special cases of our solution are exactly the same as some
solutions discovered previously. In the previous work [6] we have
applied the asymptotic conditions to derive the Schwarzschild
metric in the expanding universe, which is exactly the same as
that derived by McVittie by solving the full Einstein equations.
That demonstrates the power of this simple and straight-forward
approach. In this work we follow the same procedure to derive the
metric for the Reissner-Nordstr$\ddot{o}$m Black Holes in
Friedman-Robertson-Walker Universe. We then study the influences
of the evolution of the universe on the size of the black hole.
Finally in order to study the motion of the planet, we rewrite the
metric from the cosmic coordinates system to the Schwarzschild coordinates system.\\
\section{Derivation of the metric}
\hspace*{7.5mm}The metric of Reissner-Nordstr$\ddot{o}$m black
hole in the Schwarzschild coordinates system is given by
\begin{equation}
d\tilde{s}^2=-\left(1-\frac{2M}{\tilde{r}}+\frac{Q^2}{\tilde{r}^2}\right)
d\tilde{t}^2+\left(1-\frac{2M}{\tilde{r}}+\frac{Q^2}{\tilde{r}^2}\right)^{-1}d\tilde{r}^2+\tilde{r}^2\left(d
\theta^2+\sin ^2 \theta d \phi^2\right),
\end{equation}
where $M$ and $Q$ are the mass and charge of the black hole,
respectively.\\
\hspace*{7.5mm}For our purpose we rewrite the metric Eq.(1)in the
isotropic spherical
 coordinates. We assume that $x^0=v$ and
$x^1=x$. So make variables transformation
 \begin{equation}
\tilde{t}=2v,\ \ \ \ \tilde{s}=2l,\ \ \ \
2\tilde{r}={x\left(1+\frac{M}{x}\right)^2-\frac{Q^2}{x^2}},
\end{equation}
 then we can rewrite Eq.(1) as follows
\begin{equation}
dl^2=-\frac{\left(1-\frac{M^2}{x^2}+\frac{Q^2}{x^2}\right)^2}{\left[\left(1+\frac{M}{x}\right)^2
-\frac{Q^2}{x^2}\right]^2}dv^2+\left[\left(1+\frac{M}{x}\right)^2
-\frac{Q^2}{x^2}\right]^2\left(dx^2+x^2d\theta^2+x^2\sin ^2 \theta
d \phi^2\right).
\end{equation}
As is known, the metric for the FRW (Friedman-Robertson-Walker)
universe is  given by
\begin{equation}
 dl^2=-dv^2+\frac{a^2\left(v\right)}{\left(1+kx^2/4\right)^2}\left(dx^2+x^2d\theta^2+x^2\sin^2\theta d\phi^2 \right),
\end{equation}
where $a(v)$ is the scale factor of the universe and $k$ gives the
curvature of space-time as a whole.\\
\hspace*{7.5mm} Taking account of equations Eq.(3) and Eq.(4), we
set the metric for a Reissner-Nordstr$\ddot{o}$m black hole
embedded in the FRW universe as follows
 \begin{equation}
dl^{2} = -A^{2}\left( {v, x} \right)dv^{2} + B^{2}\left( {v, x}
\right)\left(dx^2+x^2d\theta^2+x^2\sin^2\theta d\phi^2 \right).
\end{equation}
Then from equation $G_{01}=0$ one obtains
\begin{eqnarray}
A\left(v, x\right)=f\left(v\right)\frac{\dot{B}}{2B},
 \end{eqnarray}
where $"\cdot¡±"$ denotes the derivative with respect to $v$.\\
\hspace*{7.5mm}Compare the ${g_{11}}$ terms in Eq.(3) and Eq.(5),
the possible form for the function $B(v,x)$ is
\begin{equation}
B\left(v,
x\right)=\left[w\left(v,x\right)+\frac{q\left(v\right)}{x}\right]^2-\frac{s\left(v\right)}{x^2}.
\end{equation}
We note that the mass and charge of the black hole is concentrated
in the singularity. In other words, there is no space distribution
for mass and charge. Thus $q$ and $s$ which are related to the
mass and charge, respectively,  are only the
functions of time $v$.\\
\hspace*{7.5mm}Inserting Eq.(7) into Eq.(6), we obtain
\begin{equation}
A\left(v,x\right)=\frac{\frac{\dot{w}f}{w}+\left(w\dot{q}+\dot{w}q\right)\frac{f}{w^2x}
+\frac{q\dot{q}f}{w^2x^2}-\frac{\dot{s}f}{2w^2x^2}}{\left(1+\frac{q}{wx}\right)^2-\frac{s}{w^2x^2}}.
\end{equation}
In the case of $v=const$ and the asymptotically flat conditions,
$A(v=const, x)$ should be reduced to the $\sqrt{-g_{00}}$ term in
Eq.(3). Thus comparing Eq.(8) with the $\sqrt{-g_{00}}$ term in
Eq.(3), we infer the following identities should always hold
\begin{eqnarray}
\frac{{\dot{w}f}}{w}&=&1,\nonumber
\\\left(w\dot{q}+\dot{w}q\right)\frac{f}{w^2x}&=0&,\nonumber \\\frac{q\dot{q}f}{w^2x^2}&=&-\left(\frac{q}{wx}\right)^2,\nonumber \\
-\frac{\dot{s}f}{2w^2x^2}&=&\frac{s}{w^2x^2},
\end{eqnarray}
namely
\begin{equation}
{{\dot{w}f}}=w,\ \ \ \ {\dot{q}f}=-q,\ \ \ \ {\dot{s}f}=-2{s},
\end{equation}
 From Eqs.(10) we obtain
\begin{equation}
w=\frac{b\left(v\right)}{\sqrt{1+kx^2/4 }},\ \ \ \
f=\frac{b}{\dot{b}},\ \ \ \ q=\frac{M}{b},\ \ \ \
s=\frac{Q^2}{b^2},
\end{equation}
where $M$ and $Q$ are two integration constant which are related
to the mass and charge of the black hole; $b(v)$ is an arbitrary
function which is related to the scale factor of the universe; the
form of $\omega$ is obtained by inspecting the McVittie's
solution,
\begin{eqnarray}
dl^2 &=&
-\frac{\left(\frac{\sqrt{a}}{\sqrt{1+kx^2/4}}-\frac{M/\sqrt{a}}{x}\right)^2}
{\left(\frac{\sqrt{a}}{\sqrt{1+kx^2/4}}+\frac{M/\sqrt{a}}{x}
\right)^2}dv^2
+\left(\frac{\sqrt{a}}{\sqrt{1+kx^2/4}}+\frac{M/\sqrt{a}}{x}\right)^4 \nonumber\\
&& \cdot\left(dx^2+x^2d\theta^2+x^2\sin ^2 \theta d \phi^2\right),
\end{eqnarray}
where $a=a\left(v\right)$. \\
\hspace*{7.5mm}Substituting Eqs.(7, 8, 11) in Eq.(5) , we obtain
our final metric for the Reissner-Nordstr$\ddot{o}$m black hole in
the background of FRW universe
\begin{eqnarray}
dl^2 &=&
-\frac{\left[1-\frac{M^2}{a^2x^2}\left(1+kx^2/4\right)\right]^2+\frac{Q^2}{a^2x^2}\left(1+kx^2/4\right)}{\left[\left(1+\frac{M}{ax}
\sqrt{1+kx^2/4}\right)^2
-\frac{Q^2}{a^2x^2}\left(1+kx^2/4\right)\right]^2}dv^2\nonumber\\&
&
+\frac{a^2}{\left(1+kx^2/4\right)^2}\left[\left(1+\frac{M}{ax}\sqrt{1+kx^2/4}\right)^2
-\frac{Q^2}{a^2x^2}\left(1+kx^2/4\right)\right]^2\nonumber\\&
&\cdot\left(dx^2+x^2d\theta^2+x^2\sin ^2 \theta d \phi^2\right),
\end{eqnarray}
where we have made a variable replacement $b(v)^2\rightarrow
a(v)$. Eq.(13) is derived from Eq.(7) and when $a(v)=constant,k=0$
Eq.(13) restores Eq.(3). So Eq.(3) satisfies Eq.(7).\\
\hspace*{7.5mm}For the Reissner-Nordstr$\ddot{o}$m-de Sitter
metric, we have $a=e^{Ht},k=0$, so Eq.(13) becomes
\begin{eqnarray} dl^2 =
-\frac{\left[1-\frac{M^2}{a^2x^2}+\frac{Q^2}{a^2x^2}\right]^2}{\left[\left(1+\frac{M}{ax}
\right)^2 -\frac{Q^2}{a^2x^2}\right]^2}dv^2
+{a^2}\left[\left(1+\frac{M}{ax}\right)^2
-\frac{Q^2}{a^2x^2}\right]^2\left(dx^2+x^2d\theta^2+x^2\sin ^2
\theta d \phi^2\right).
\end{eqnarray}
We will show in the next section that Eq.(14) can be reduced to
the familiar form in Schwarzschild coordinates. When $H=0$,
Eq.(14) restores the Reissner-Nordstr$\ddot{o}$m metric as given
by Eq.(3). When $Q=0$, Eq.(13) becomes
\begin{eqnarray}
dl^2 &=&
-\frac{\left(1-\frac{M}{ax}\sqrt{1+kx^2/4}\right)^2}{\left(1+\frac{M}{ax}
\sqrt{1+kx^2/4}\right)^2}dv^2
+\frac{a^2}{\left(1+kx^2/4\right)^2}\left(1+\frac{M}{ax}\sqrt{1+kx^2/4}\right)^4
\nonumber\\ && \cdot\left(dx^2+x^2d\theta^2+x^2\sin ^2 \theta d
\phi^2\right),
\end{eqnarray}
It is just the McVittie solution. Another special case of our
solution is for the extreme Reissner-Nordstr$\ddot{o}$m black
hole, $M=Q$, in the de Sitter universe. In this case,
 Eq.(14) is reduced to a special case of the Kastor and Traschen solution
 [2] for a single black hole,
\begin{eqnarray}
dl^2 =
-\frac{1}{\left(1+\frac{2M}{ax} \right)^2 }dv^2
+{a^2}\left(1+\frac{2M}{ax}
\right)^2\left(dx^2+x^2d\theta^2+x^2\sin ^2 \theta d
\phi^2\right).
\end{eqnarray}
In Eq.(16), $a=e^{Ht}$. It describes one extreme
Reissner-Nordstr$\ddot{o}$m black hole in the de Sitter universe.
\section{further discussion on the metric}
\hspace*{7.5mm}In this section we will verify that Eq.(13)
satisfies Einstein-Maxwell equations. The Einstein-Maxwell
equations may be written as
\begin{eqnarray}
G_{\mu\nu}&=&8 \pi \left(T_{\mu\nu}+E_{\mu\nu}\right), \nonumber
\\ F_{\mu\nu}&=& A_{\mu;\nu}-A_{\nu;\mu},\nonumber \\  F^{\mu\nu}_{;\nu}&=&0,
\end{eqnarray}
where $T_{\mu\nu}$ and $E_{\mu\nu}$ are the energy momentum for
the perfect fluid and electromagnetic fields, respectively, which
are defined by
\begin{eqnarray}
T_{\mu\nu}&=&\left(\rho+p\right)U_{\mu}U_{\nu}+pg_{\mu\nu},
\nonumber \\
E_{\mu\nu}&=&\frac{1}{4\pi}\left(F_{\mu\alpha}F^{\alpha}_{\nu}-\frac{1}{4}g_{\mu\nu}F_{\alpha\beta}F^{\alpha\beta}\right),
\end{eqnarray}
where $\rho$ and $p$ are the energy density and pressure.
$U_{\mu}$ is the 4-velocity of the particles. $F_{\mu\nu}$ and
$A_{\mu}$ are the tensor and the potential
for electromagnetic fields.\\
\hspace*{7.5mm}Input the components of the metric, Eq.(13), to the
Maple software package, we obtain the Einstein tensor $G_{\mu\nu}$
and the energy momentum tensor $T_{\mu\nu}$ and $E_{\mu\nu}$ for
the perfect fluid and the electromagnetic fields
\begin{eqnarray}
T_0^0&=&\rho, \ \ \ \ \ \ \ \ \ \ \ \ T_1^1=T_2^2=T_3^3=p,\nonumber\\
2\pi E_0^0&=&2\pi
E_1^1=\frac{Q^2\left(1+kx^2/4\right)}{x^4a^4\left[\left(1+\frac{M}{ax}\sqrt{1+kx^2/4}\right)^2
-\frac{Q^2}{a^2x^2}\left(1+kx^2/4\right)\right]^4},\nonumber\\
2\pi E_2^2&=&2\pi
E_3^3=-\frac{Q^2\left(1+kx^2/4\right)}{x^4a^4\left[\left(1+\frac{M}{ax}\sqrt{1+kx^2/4}\right)^2
-\frac{Q^2}{a^2x^2}\left(1+kx^2/4\right)\right]^4}.
\end{eqnarray}
Substituting the above components of electromagnetic tensor in the
second equation of Eqs.(18), we obtain the non-vanishing
components of electromagnetic tensor $F_{\mu\nu}$
\begin{eqnarray}
F^{01}&=&\frac{\left(1+kx^2/4\right)^{3/2}Q}{x^2a^3\left[1-\frac{M^2}{a^2x^2}\left(1+kx^2/4\right)+\frac{Q^2}{a^2x^2}
\left(1+kx^2/4\right)\right]}\nonumber\\
&&
\cdot\frac{1}{\left[\left(1+\frac{M}{ax}\sqrt{1+kx^2/4}\right)^2
-\frac{Q^2}{a^2x^2}\left(1+kx^2/4\right)\right]^2}.
\end{eqnarray}
Then substituting Eq.(20) in the second equation of Eqs.(17), we
obtain the non-vanishing components of the potential $A_{\mu}$
\begin{equation}
A_{0}=\int F^{01}g_{00}g_{11}dx.
\end{equation}
It is a straightforward work to verify that Eq.(20) also satisfies
the last equation in Eqs.(17). Since $G^{\mu\nu}_{;\nu}=0$ always
holds, thus from Einstein equations we have
$0=T^{\mu{\nu}}_{;\nu}+E^{\mu{\nu}}_{;\nu}$. On the other hand, we
have the relation \begin{eqnarray} &&4\pi
E^{\mu{\nu}}_{;\nu}=F^{\mu\alpha}F^{\nu}_{\alpha;\nu}+F^{\mu\alpha}_{;\nu}
F^{\nu}_{\alpha}-\frac{1}{2}g^{\mu\nu}F^{\alpha\beta}F_{\alpha\beta}\nonumber
\\&& =
F^{\mu\alpha}F^{\nu}_{\alpha;\nu}+\frac{1}{2}g^{\mu\rho}F^{\nu\sigma}
\left(F_{\rho\sigma;\nu}-F_{\rho\nu;\sigma}-F_{\sigma\nu\rho}
\right)\nonumber\\ && =-F^{\mu\alpha}J_{\alpha}=0.
\end{eqnarray}
So both $T_{\mu\nu}$ and $E_{\mu\nu}$ satisfy Bianchi identity.
We therefore conclude that Eq.(13) is an exact solution of the Einstein-Maxwell equations.\\
\hspace*{7.5mm}To show how the two parameters $M$ and $Q$ are
related to the mass and charge of the black hole, we assume the
evolution of the universe is much slower and approximately adopt
the mass formula for the stationary space-time [8](To our
knowledge, we can only define the mass for the stationary and
asymptotically flat space-time). We find the mass $M_{0}$ and
charge $Q_{0}$ of the black hole are given by
\begin{eqnarray}
&&M_{0}\equiv-\frac{1}{8\pi}\int_{S}\epsilon_{abcd}\nabla^{c}\xi^{d}=\frac{M}{a},\nonumber
\\ && Q_{0}\equiv \frac{1}{4\pi}\int_{S}\epsilon_{abcd}
F^{cd}=\frac{Q}{a}.
\end{eqnarray}
Thus for the observer in the infinity, the black hole's mass and
charge will decrease with the expansion of the universe and
increase with the contraction of the universe. We will return to
this point in the following discussion again.\\
\hspace*{7.5mm}Let's now consider the two typical surfaces of the
black hole in the cosmic coordinates system. We obtain the radius
of the time-like limit surface(TLS) [8]
\begin{eqnarray}
 r_{TLS}=\sqrt{\frac{M^2}{a^2}-\frac{Q^2}{a^2}},
 \end{eqnarray}
 and the time derivative of the radius of the event horizon [8]
\begin{eqnarray}
 \dot{r}_{EH}=\pm\frac{1-\frac{M^2}{a^2r_{EH}^2}+\frac{Q^2}{a^2r_{EH}^2}}{a \left[\left(1+\frac{M}{ar_{EH}}
 \right)^2-\frac{Q^2}{a^2r_{EH}^2}\right]^2},
 \end{eqnarray}
where two signs $"+"$ and $"-"$ correspond to the expanding and
the contracting universe, respectively. Since the event horizon is
always in the inner of TLS, we have $\dot{r}_{EH}<0$ for expanding
universe  and $\dot{r}_{EH}>0$ for contracting universe.\\
\hspace*{7.5mm}Eqs.(24, 25) tell us the typical scales of the
black hole are closely related to the evolution of the universe.
They shrink with the expansion of the universe and expand with the
contracting of the universe. For asymptotically flat background,
$a=1$ and $\dot{r}_{EH}=0$, these two kind of surfaces coincide
\begin{eqnarray}
 r_{TLS}=r_{EH}=\sqrt{{M^2}-{Q^2}}.
 \end{eqnarray}
Comparing Eq.(26) with Eq.(24), we find that the black hole in the
expanding universe has the mass $M_{0}=M/a$ and charge $Q_{0}/a$.
They are both dependent
on the scale of the universe and not a constant.\\
\hspace*{7.5mm}From Eq.(23), one obtain the lifetime of a black
hole
\begin{eqnarray}
 \tau\simeq\frac{1}{H}.
 \end{eqnarray}
It is approximately the age of the universe.
\section{equation of motion of a planet}
\hspace*{7.5mm}In this section, we turn to the discussion of the
problem of the orbit for a planet in the expanding universe.
Current measurements of the microwave background radiation show
that our universe is highly likely flat in space [7]. So in the
next we consider the local dynamics in the space-flat space case,
ie., $k=0$. Eq.(13) is then reduced to
\begin{equation}
dl^2=-\frac{\left(1-\frac{M^2}{a^2x^2}+\frac{Q^2}{a^2x^2}\right)^2}{\left[\left(1+\frac{M}{ax}\right)^2
-\frac{Q^2}{a^2x^2}\right]^2}dv^2+a^2\left[\left(1+\frac{M}{ax}\right)^2
-\frac{Q^2}{a^2x^2}\right]^2\left(dx^2+x^2d\theta^2+x^2\sin ^2
\theta d \phi^2\right).
\end{equation}
\hspace*{7.5mm}In order to study the motion of a planet, we should
rewrite Eq.(28) in the Schwarzschild or solar coordinates system.
Similar to Eq.(2), make variables transformation as follows
 \begin{equation}
{T}=2v,\ \ \ \ {s}=2l,\ \ \ \
2{r}={ax\left(1+\frac{M}{ax}\right)^2-\frac{Q^2}{a^2x^2}},
\end{equation}
Then Eq.(28) becomes
\begin{eqnarray}
ds^2&=&-\left(1-\frac{2M}{r}+\frac{Q^2}{r^2}-H^2r^2\right)dT^2+(1-\frac{2M}{r}+\frac{Q^2}{r^2})^{-1}dr^2\nonumber\\&
&-2rH\left(1-\frac{2M}{r}+\frac{Q^2}{r^2}\right)^{-1/2}dTdr
+r^2\left(d\theta^2+\sin^2\theta d\phi^2\right),
\end{eqnarray}
where $H\equiv \frac{1}{a}\frac{da}{dT}$ is the Hubble parameter.
 The coordinate system of
$(T,r,\theta,\phi)$ is not orthogonal. We can eliminate the
coefficient of $dTdr$ by introducing a new time coordinate, $t$.
The form of Eq.(30) suggests we set
\begin{equation}
dt=F\left(T,r\right)dT+{F\left(T,r\right)}{rH\left(1-\frac{2M}{r}+\frac{Q^2}{r^2}\right)^{-1/2}}{\left(1-\frac{2M}{r}+\frac{Q^2}{r^2}-H^2r^2\right)^{-1}}dr,
\end{equation}
where $F(T,r)$ is a perfect differential factor and it always
exists. Then Eq.(29) can be written as
\begin{equation}
ds^2=-\left(1-\frac{2M}{r}+\frac{Q^2}{r^2}-H^2r^2\right)F^2dt^2+\left(1-\frac{2M}{r}+\frac{Q^2}{r^2}-H^2r^2\right)^{-1}dr^2
+r^2d\Omega^2,
\end{equation}
where $H$ and $F$ are both the functions of variables $t$ and $r$.
If $\rho=p=0$, we have $H=0$. Eq.(30) and Eq.(32) both turn into
the static Reissner-Nordstr$\ddot{o}$m solution. If
$2M/r\longrightarrow 0$ and $Q^2/r^2\longrightarrow 0$, Eq.(32)
represents the solution for the Friedman-Robertson-Walker universe
in Schwarzschild or solar coordinate system. For the de Sitter
universe, $H$ is a constant. Eq.(31) tells us we may choose
$F(t,r)=1$. Then Eq.(32) is just the well-known
Reissner-Nordstr$\ddot{o}$m-de Sitter space-time. It is a static
space-time. Thus the geodesic is time-independent. In other words,
the orbit of the planet does
not vary in the background of de Sitter universe.\\
\hspace*{7.5mm}In general, $H$ is the function of  $t$ and $r$ .
So Eq.(32) is a non-stationary space-time and the geodesic in this
space-time is generally time-dependent. Thus we conclude that
universe expansion would influence the orbit of planet. However,
the effect term $H^2r^2$ is very small. Since
\begin{equation}
\varepsilon\equiv\frac{H^2r^2}{2M/r}=\frac{\rho}{M/(4\pi
r^3/3)}=\frac{\rho}{\tilde{\rho}},
\end{equation}
where $\rho$ is the universe energy density and $\tilde{\rho}$ is
the energy density within the region $r$ where the energy $M$ is
distributed. The ratio of $\rho$ to $\tilde{\rho}$ varies from
$4\times10^{-34}$ for Mercury-Sun system (In detail, $r$
represents the distance from the Sun to Mercury and $M$ is the
mass of the Sun. The mass of the planet is much smaller than the
Sun.), $1.8\times10^{-28}$ for Neptune-Sun system to
$10^{-7}$ for Galaxy.\\
\hspace*{7.5mm}Keep this in mind or regard $H$ as a constant, we
obtain the motion equation of a planet
\begin{equation}
u''+u=\frac{M}{L^2}-\frac{Q^2u}{L^2}+3Mu^2-2Q^2u^3-\frac{H^2}{L^2u^3},
\end{equation}
where $u\equiv 1/r$ and the prime denotes differentiation with
respect to $\phi$. $L$ is the angular momentum of the planet. The
term $M/L^2$ is the most significant one. The two terms $3Mu^2$
and $H^2/(L^2u^3)$ come from the GR (general relativity) effect
and CE (cosmic expansion) effect, respectively. $Q^2u/L^2$ and
$-2Q^2u^2$ are related to the charge of the source. Eq.(34) tells
us the orbit of the test body will be influenced by the expansion
of the universe. Since
\begin{equation}
\frac{H^2/\left(L^2u^3\right)}{M/L^2}=2\varepsilon=8\times10^{-34}\
\ $for Mercury$,\ \ 3.6\times10^{-28} \ \ $for Neptune$,
\end{equation}
so the influence is related to the ratio of $\rho$ to
$\tilde{\rho}$.  It is very small and negligible.

\section{conclusion and discussion}
\hspace*{7.5mm}In conclusion, we have presented the metric for the
Reissner-Nordstr$\ddot{o}$m black hole in the background of FRW
universe. It extends McVittie's solution,
Reissner-Nordstr$\ddot{o}$m-de Sitter solution and the special
case for a single black hole of the Kastor and Traschen solution.
Assume the evolution of the universe is much slower and adopt the
formulas for computing the mass and charge of stationary
space-time, we find that both the mass and charge of the black
hole decrease with the expansion of the universe and increase with
the contraction of the universe. We also find that the two typical
scales, the time-like surface and the event horizon, of the black
hole both shrink with the expansion of the universe and expand
with the contraction of the universe. This is due to the fact that
the mass and charge of the black hole are both varying with the
evolution
of the universe.\\
\hspace*{7.5mm}To obtain the equation of motion of a planet, we
rewrote the metric from the cosmic coordinate system to the
Schwarzschild or solar coordinate system and deduced the geodesic
equation. The equation shows that the orbit of the planet in the
Reissner-Nordstr$\ddot{o}$m field embedded in the FRW universe
will be influenced by the evolution of the universe. The magnitude
of influence depends on the ratio $\varepsilon$ between the energy
density of the system and the energy density of the universe.
Since the ratio $\varepsilon$ is extremely small, varying from
$4\times10^{-34}$ for Mercury-Sun system, $1.8\times10^{-28}$ for
Neptune-Sun system to $10^{-7}$ for Galaxy, the influence of
the expansion of the universe is very small and negligible [9].\\
\hspace*{7.5mm}\acknowledgements We thank the anonymous referee
for the expert and insightful comments, which have certainly
improved the paper significantly. This study is supported in part
by the Special Funds for Major State Basic Research Projects and
by the National Natural Science Foundation of China. SNZ also
acknowledges supports by NASA's Marshall Space Flight Center and
through NASA's Long Term Space Astrophysics Program.


\begin{references}
\bibitem{1}McVittie G C 1933 Mon. Not. R. Astron. Soc. {\bf 93} 325
\bibitem{2}Kastor D and Traschen J 1993 Phys. Rev. D {\bf 47}
5401
\bibitem{3}Shiromizu T and Gen U 2000 Class. Quntum. Grav. {\bf 17}
1361
\bibitem{4}Nayak K R, MacCallum M A H and Vishveshvara C V 2000 Phys. Rev. D {\bf 63}
024020
\bibitem{5}Nayak K R and Vishveshvara C V 2000 "Geometry of the Kerr Black Hole in
the Einstein Cosmological Background," report.
\bibitem{6}Gao C J and Zhang S N "Schwarzschild Metric in the Expanding Universe", submitted to Phys. Rev. D.
\bibitem{8}Hanany S et al. 2000 Astrophys. J. {\bf 545} L5
\bibitem{9}Wald R M 1984 General relativity (The University of Chicago Press)
\bibitem{10}P. D. Noerdlinger and V. Petrosion, Astrophys. J. {\bf 168}, 1(1971).
\end{references}
\end{document}